\documentclass[conference,a4paper]{IEEEtran}

\usepackage[left=0.7in, right=0.7in, bottom=1in, top=0.75in]{geometry}

\usepackage{multicol}
\usepackage{graphicx}
\graphicspath{{Figures/}}
\usepackage{epstopdf}
\usepackage{acronym}
\usepackage{amsmath}
\usepackage{color, colortbl}
\usepackage{subfigure}
\usepackage[english]{babel}
\usepackage{blindtext}
\usepackage{algorithm} 
\usepackage{algorithmic} 
\usepackage{multirow}
\usepackage{color}
\usepackage{subfigure}
\usepackage{balance}
\usepackage{caption}

\begin{document}

\newacro{3GPP}{third generation partnership project}
\newacro{4G}{4{th} generation}
\newacro{5G}{5{th} generation}

\newacro{ADC}{analogue-to-digital conversion}
\newacro{AED}{accumulated euclidean distance}
\newacro{AGC}{automatic gain control}
\newacro{AI}{artificial intelligence}
\newacro{AN}{artificial noise}
\newacro{ASE}{amplified spontaneous emission}
\newacro{ASIC}{application specific integrated circuit}
\newacro{AWG}{arbitrary waveform generator}
\newacro{AWGN}{additive white Gaussian noise}
\newacro{A/D}{analog-to-digital}

\newacro{B2B}{back-to-back}
\newacro{BCF}{bandwidth compression factor}
\newacro{BCJR}{Bahl-Cocke-Jelinek-Raviv}
\newacro{BDM}{bit division multiplexing}
\newacro{BED}{block efficient detector}
\newacro{BER}{bit error rate}
\newacro{Block-SEFDM}{block-spectrally efficient frequency division multiplexing}
\newacro{BLER}{block error rate}
\newacro{BPSK}{binary phase shift keying}
\newacro{BS}{base station}
\newacro{BSS}{best solution selector}
\newacro{BU}{butterfly unit}

\newacro{CapEx}{capital expenditure}
\newacro{CA}{carrier aggregation}
\newacro{CBS}{central base station}
\newacro{CC}{component carriers}
\newacro{CCDF}{complementary cumulative distribution function}
\newacro{CCs}{component carriers}
\newacro{CD}{chromatic dispersion}
\newacro{CDF}{cumulative distribution function}
\newacro{CDI}{channel distortion information}
\newacro{CDMA}{code division multiple access}
\newacro{CI}{constructive interference}
\newacro{CIR}{carrier-to-interference power ratio}
\newacro{CMOS}{complementary metal-oxide-semiconductor}
\newacro{CNN}{convolutional neural network}
\newacro{CoMP}{coordinated multiple point}
\newacro{CO-SEFDM}{coherent optical-SEFDM}
\newacro{CP}{cyclic prefix}
\newacro{CPE}{common phase error}
\newacro{CRVD}{conventional real valued decomposition}
\newacro{CR}{cognitive radio}
\newacro{CRC}{cyclic redundancy check}
\newacro{CS}{central station}
\newacro{CSI}{channel state information}
\newacro{CSIT}{channel state information at transmitter}
\newacro{CSPR}{carrier to signal power ratio}
\newacro{CWT}{continuous wavelet transform}
\newacro{C-RAN}{cloud-radio access networks}

\newacro{DAC}{digital-to-analogue conversion}
\newacro{DBP}{digital backward propagation}
\newacro{DC}{direct current}
\newacro{DCT}{discrete cosine transform}
\newacro{DDC}{digital down-conversion}
\newacro{DDO-OFDM}{directed detection optical-OFDM}
\newacro{DDO-OFDM}{direct detection optical-OFDM}
\newacro{DDO-SEFDM}{directed detection optical-SEFDM}
\newacro{DFB}{distributed feedback}
\newacro{DFDMA}{distributed FDMA}
\newacro{DFT}{discrete Fourier transform}
\newacro{DFrFT}{discrete fractional Fourier transform}
\newacro{DMA}{direct memory access}
\newacro{DMRS}{demodulation reference signal}
\newacro{DOFDM}{dense orthogonal frequency division multiplexing}
\newacro{DP}{dual polarization}
\newacro{DPC}{dirty paper coding}
\newacro{DSB}{double sideband}
\newacro{DSL}{digital subscriber line}
\newacro{DSP}{digital signal processors}
\newacro{DVB}{digital video broadcast}
\newacro{DWT}{discrete wavelet transform}
\newacro{D/A}{digital-to-analog}

\newacro{ECC}{error correcting codes}
\newacro{ECL}{external-cavity laser}
\newacro{ECOC}{error-correcting output codes}
\newacro{EDFA}{erbium doped fiber amplifier}
\newacro{EE}{energy efficiency}
\newacro{eMBB}{enhanced mobile broadband}
\newacro{eNB-IoT}{enhanced NB-IoT}
\newacro{EPA}{extended pedestrian A}
\newacro{EVM}{error vector magnitude}

\newacro{Fast-OFDM}{fast-orthogonal frequency division multiplexing}
\newacro{FBMC}{filterbank based multicarrier }
\newacro{FCE}{full channel estimation}
\newacro{FD}{fixed detector}
\newacro{FDD}{frequency division duplexing}
\newacro{FDM}{frequency division multiplexing}
\newacro{FDMA}{frequency division multiple access}
\newacro{FE}{full expansion}
\newacro{FEC}{forward error correction}
\newacro{FEXT}{far-end crosstalk}
\newacro{FF}{flip-flop}
\newacro{FFT}{fast Fourier transform}
\newacro{FFTW}{fastest Fourier transform in the west}
\newacro{FIFO}{first in first out}
\newacro{F-OFDM}{filtered-orthogonal frequency division multiplexing}
\newacro{FPGA}{field programmable gate array}
\newacro{FrFT}{fractional Fourier transform}
\newacro{FSD}{fixed sphere decoding}
\newacro{FSD-MNSF}{FSD-modified-non-sort-free}
\newacro{FSD-NSF}{FSD-non-sort-free}
\newacro{FSD-SF}{FSD-sort-free}
\newacro{FSK}{frequency shift keying}
\newacro{FTN}{faster than Nyquist}
\newacro{FTTB}{fiber to the building}
\newacro{FTTC}{fiber to the cabinet}
\newacro{FTTdp}{fiber to the distribution point}
\newacro{FTTH}{fiber to the home}

\newacro{GB}{guard band}
\newacro{GFDM}{generalized frequency division multiplexing}
\newacro{GPU}{graphics processing unit}
\newacro{GSM}{global system for mobile communication}
\newacro{GUI}{graphical user interface}

\newacro{HC-MCM}{high compaction multi-carrier communication}
\newacro{HPA}{high power amplifier}

\newacro{IC}{integrated circuit}
\newacro{ICI}{inter carrier interference}
\newacro{ID}{iterative detection}
\newacro{IDCT}{inverse discrete cosine transform}
\newacro{IDFT}{inverse discrete Fourier transform}
\newacro{IDFrFT}{inverse discrete fractional Fourier transform}
\newacro{ID-FSD}{iterative detection-FSD}
\newacro{ID-SD}{ID-sphere decoding}
\newacro{IF}{intermediate frequency}
\newacro{IFFT}{inverse fast Fourier transform}
\newacro{IFrFT}{inverse fractional Fourier transform}
\newacro{IMD}{intermodulation distortion}
\newacro{IoT}{internet of things}
\newacro{IOTA}{isotropic orthogonal transform algorithm}
\newacro{IP}{intellectual property}
\newacro{ISC}{interference self cancellation}
\newacro{ISI}{inter symbol interference}
\newacro{ISM}{industrial, scientific and medical}

\newacro{KNN}{k-nearest neighbours}

\newacro{LDPC}{low density parity check}
\newacro{LFDMA}{localized FDMA}
\newacro{LLR}{log-likelihood ratio}
\newacro{LNA}{low noise amplifier}
\newacro{LO}{local oscillator}
\newacro{LOS}{line-of-sight}
\newacro{LPWAN}{low power wide area network}
\newacro{LS}{least square}
\newacro{LTE}{long term evolution}
\newacro{LTE-Advanced}{long term evolution-advanced}
\newacro{LUT}{look-up table}

\newacro{MA}{multiple access}
\newacro{MAC}{media access control}
\newacro{MASK}{m-ary amplitude shift keying}
\newacro{MCM}{multi-carrier modulation}
\newacro{MC-CDMA}{multi-carrier code division multiple access}
\newacro{MCS}{modulation and coding scheme}
\newacro{MF}{matched filter}
\newacro{MIMO}{multiple input multiple output}
\newacro{ML}{maximum likelihood}
\newacro{MLSD}{maximum likelihood sequence detection}
\newacro{MMF}{multi-mode fiber}
\newacro{MMSE}{minimum mean squared error}
\newacro{mMTC}{massive machine-type communication}
\newacro{MNSF}{modified-non-sort-free}
\newacro{MOFDM}{masked-OFDM}
\newacro{MRVD}{modified real valued decomposition}
\newacro{MS}{mobile station}
\newacro{MSE}{mean squared error}
\newacro{MTC}{machine-type communication}
\newacro{MUSA}{multi-user shared access}
\newacro{MU-MIMO}{multi-user multiple-input multiple-output}
\newacro{MZM}{Mach-Zehnder modulator}
\newacro{M2M}{machine to machine}

\newacro{NB-IoT}{narrowband IoT}
\newacro{NEXT}{near-end crosstalk}
\newacro{NG-IoT}{next generation IoT}
\newacro{NLOS}{non-line-of-sight}
\newacro{NN}{neural network}
\newacro{NOFDM}{non-orthogonal frequency division multiplexing}
\newacro{NOMA}{non-orthogonal multiple access}
\newacro{NoFDMA}{non-orthogonal frequency division multiple access}
\newacro{NP}{non-polynomial}
\newacro{NR}{new radio}
\newacro{NSF}{non-sort-free}
\newacro{NWDM}{Nyquist wavelength division multiplexing }
\newacro{Nyquist-SEFDM}{Nyquist-spectrally efficient frequency division multiplexing}

\newacro{OBM-OFDM}{orthogonal band multiplexed OFDM}
\newacro{OF}{optical filter}
\newacro{OFDM}{orthogonal frequency division multiplexing}
\newacro{OFDMA}{orthogonal frequency division multiple access}
\newacro{OMA}{orthogonal multiple access}
\newacro{OpEx}{operating expenditure}
\newacro{OQAM}{offset-QAM}
\newacro{OSI}{open systems interconnection}
\newacro{OSNR}{optical signal-to-noise ratio}
\newacro{OSSB}{optical single sideband}
\newacro{OTA}{over-the-air}
\newacro{Ov-FDM}{Overlapped FDM}
\newacro{O-SEFDM}{optical-spectrally efficient frequency division multiplexing}
\newacro{O-FOFDM}{optical-fast orthogonal frequency division multiplexing}
\newacro{O-OFDM}{optical-orthogonal frequency division multiplexing}

\newacro{PA}{power amplifier}
\newacro{PAPR}{peak-to-average power ratio}
\newacro{PCE}{partial channel estimation}
\newacro{PD}{photodiode}
\newacro{PDF}{probability density function}
\newacro{PDP}{power delay profile}
\newacro{PDMA}{polarisation division multiple access}
\newacro{PDM-OFDM}{polarization-division multiplexing-OFDM}
\newacro{PDM-SEFDM}{polarization-division multiplexing-SEFDM}
\newacro{PDSCH}{physical downlink shared channel}
\newacro{PE}{processing element}
\newacro{PED}{partial Euclidean distance}
\newacro{PLS}{physical layer security}
\newacro{PMD}{polarization mode dispersion}
\newacro{PON}{passive optical network}
\newacro{PPM}{parts per million}
\newacro{PRB}{physical resource block}
\newacro{PSD}{power spectral density}
\newacro{PSK}{pre-shared key}

\newacro{PU}{primary user}
\newacro{PXI}{PCI extensions for instrumentation}
\newacro{P/S}{parallel-to-serial}

\newacro{QAM}{quadrature amplitude modulation}
\newacro{QoS}{quality of service}
\newacro{QPSK}{quadrature phase-shift keying}

\newacro{RAUs}{remote antenna units}
\newacro{RBW}{resolution bandwidth}
\newacro{RF}{radio frequency}
\newacro{RMS}{root mean square}
\newacro{RoF}{radio-over-fiber}
\newacro{ROM}{read only memory}
\newacro{RRC}{root raised cosine}
\newacro{RSC}{recursive systematic convolutional}
\newacro{RTL}{register transfer level}
\newacro{RVD}{real valued decomposition}

\newacro{ScIR}{sub-carrier to interference ratio}
\newacro{SCMA}{sparse code multiple access}
\newacro{SC-FDMA}{single carrier frequency division multiple access}
\newacro{SC-SEFDMA}{single carrier spectrally efficient frequency division multiple access}
\newacro{SD}{sphere decoding}
\newacro{SDP}{semidefinite programming}
\newacro{SDR}{software-defined radio}
\newacro{SE}{spectral efficiency}
\newacro{SEFDM}{spectrally efficient frequency division multiplexing}
\newacro{SEFDMA}{spectrally efficient frequency division multiple access} 
\newacro{SF}{sort-free}
\newacro{SGDM}{stochastic gradient descent with momentum}
\newacro{SIC}{successive interference cancellation}
\newacro{SiGe}{silicon-germanium}
\newacro{SINR}{signal-to-interference-plus-noise ratio}
\newacro{SISO}{single-input single-output}
\newacro{SMF}{single mode fiber}
\newacro{SNR}{signal-to-noise ratio}
\newacro{SP}{shortest-path}
\newacro{SRS}{sounding reference signal}
\newacro{SSB}{single-sideband}
\newacro{SSBI}{signal-signal beat interference}
\newacro{SSMF}{standard single mode fiber}
\newacro{STBC}{space time block coding}
\newacro{STO}{symbol timing offset}
\newacro{SU}{secondary user}
\newacro{SVD}{singular value decomposition}
\newacro{SVM}{support vector machine}
\newacro{SVR}{singular value reconstruction}
\newacro{S/P}{serial-to-parallel}

\newacro{TDD}{time division duplexing}
\newacro{TDMA}{time division multiple access }
\newacro{TFP}{time frequency packing}
\newacro{THP}{Tomlinson-Harashima precoding}
\newacro{TOFDM}{truncated OFDM}
\newacro{TSVD}{truncated singular value decomposition}
\newacro{TSVD-FSD}{truncated singular value decomposition-fixed sphere decoding}

\newacro{UAV}{unmanned aerial vehicle}
\newacro{UCR}{user compression ratio}
\newacro{UFMC}{universal-filtered multi-carrier}
\newacro{URLLC}{ultra-reliable and low-latency communication}
\newacro{USRP}{universal software radio peripheral}

\newacro{VDSL}{very-high-bit-rate digital subscriber line}
\newacro{VDSL2}{very-high-bit-rate digital subscriber line 2}
\newacro{VHDL}{very high speed integrated circuit hardware description language}
\newacro{VLC}{visible light communication}
\newacro{VLSI}{very large scale integration}
\newacro{VOA}{variable optical attenuator}
\newacro{VP}{vector perturbation}
\newacro{VSSB-OFDM}{virtual single-sideband OFDM}

\newacro{WAN}{wide area network}
\newacro{WCDMA}{wideband code division multiple access}
\newacro{WDM}{wavelength division multiplexing}
\newacro{WDS}{waveform-defined security}
\newacro{WiFi}{wireless fidelity}
\newacro{WiGig}{Wireless Gigabit Alliance}
\newacro{WiMAX}{Worldwide interoperability for Microwave Access}
\newacro{WLAN}{wireless local area network}
\newacro{WSS}{wavelength selective switch}

\newacro{ZF}{zero forcing}
\newacro{ZP}{zero padding}


\title{Wavelet Classification for Over-the-Air Non-Orthogonal Waveforms }
\author{\IEEEauthorblockN{Tongyang Xu and Izzat Darwazeh}
 \IEEEauthorblockA{Department of Electronic and Electrical Engineering,
University College London, London, UK\\
 Email: {tongyang.xu.11@ucl.ac.uk, i.darwazeh@ucl.ac.uk}}}

\maketitle

\begin{abstract}

Non-cooperative communications using non-orthogonal multicarrier signals are challenging since self-created inter carrier interference (ICI) exists, which would prevent successful signal classification. Deep learning (DL) can deal with the classification task without domain-knowledge at the cost of training complexity since neural network hyperparameters have to be extensively tuned. Previous work showed that a tremendously trained convolutional neural network (CNN) classifier can efficiently identify feature-diversity dominant signals while it failed when feature-similarity dominates. Therefore, a pre-processing strategy, which can amplify signal feature diversity is of great importance. This work applies single-level wavelet transform to manually extract time-frequency features from non-orthogonal signals. Composite statistical features are investigated and the wavelet enabled two-dimensional time-frequency feature grid is further simplified into a one-dimensional feature vector via proper statistical transform. The dimensionality reduced features are fed to an error-correcting output codes (ECOC) model, consisting of multiple binary support vector machine (SVM) learners, for multiclass signal classification. Low-cost experiments reveal 100\% classification accuracy for feature-diversity dominant signals and 90\% for feature-similarity dominant signals, which is nearly 28\% accuracy improvement when compared with the CNN classification results.

\end{abstract}


\section{Introduction}

In traditional communication systems, cooperation between a transmitter and a receiver is the default configuration to ensure reliable signal recovery at the receiver. Therefore, signal format is the important side information that has to be mutually known to both transmitters and receivers. The side information is delivered from a transmitter to notify a receiver and such an overhead would additionally occupy either time, frequency or space resources. Furthermore, wireless channels are time-variant and the side information of signal format would be out of date when a signal reaches a receiver after a time delay, which would subsequently cause inaccurate signal detection. Therefore, a more reliable solution is required, which avoids transmitting side information and lets the receiver timely extract signal format information from received signals.

An intelligent receiver can automatically identify signal formats based on data training. Deep learning (DL) is initially proposed to deal with image processing since it can automatically and efficiently extract features from two-dimensional images. The representative deep learning strategy is \ac{CNN}, which employs multiple convolutional layers for feature extractions. CNN has been successfully applied in single carrier modulation classification \cite{OShea_classification_2018} and multicarrier \ac{OFDM} modulation classification \cite{OFDM_classification_Zhou2019ARM, OFDM_classification_access2019}. The classification for non-orthogonal \ac{SEFDM} signals \cite{TongyangTVT2017} has been theoretically and practically investigated in work \cite{tongyang_VTC2020_DL_classification}, in which a trained \ac{CNN} classifier can efficiently identify feature-diversity dominant signals while it cannot accurately classify feature-similarity dominant signals. Although the deep learning classifier is trained to automatically extract signal features without domain-knowledge, the tremendous fine-tuning for optimal neural network hyperparameters is time consuming and inefficient. Therefore, manually extracting signal features, based on expert knowledge and traditional machine learning (ML), would be more efficient and convincing.

This work will firstly study different statistical features in \ac{SVM} for non-orthogonal signal classification. Modelling results reveal that either time-domain or frequency-domain statistical features are unable to train accurate classifiers for non-orthogonal signals. Therefore, a wavelet transform \cite{wavelet_tour_Mallat_2008, wavelet_transform_TIT_1990} based time-frequency feature extraction approach is applied in this work. Previous work has explored multilevel structured wavelet decomposition \cite{wavelet_classification_TPAMI_1997, wavelet_classification_TITB_2007} and wavelet scattering \cite{wavelet_scattering_TSP_2014} for feature extraction and classification. This work focuses on a single-level wavelet filtering (WF) strategy. Results indicate that the time-frequency feature with statistical dimensionality reduction can assist \ac{SVM} to identify signals at high accuracy. In addition, this work evaluates classifier accuracy for signals at different Es/N0. Finally, a low-cost experiment is set up to verify the trained classifiers using over-the-air signals.

The main contributions of this work are as the following.
\begin{itemize}
\item{ Statistical features are investigated in \ac{SVM} for non-orthogonal signal classification.  } 
\item{ Two-dimensional time-frequency features are evaluated via single-level wavelet transform. Various time-frequency feature dimensionality reduction methods are studied to simplify the features and further improve the classification accuracy.  } 
\item{ Low-cost over-the-air experiment is designed for non-orthogonal signal classification. Practical results verify the robustness of the wavelet classification.} 
\end{itemize}

\section{SEFDM Waveform}

\begin{figure}[t!]
\begin{center}
\includegraphics[scale=0.35]{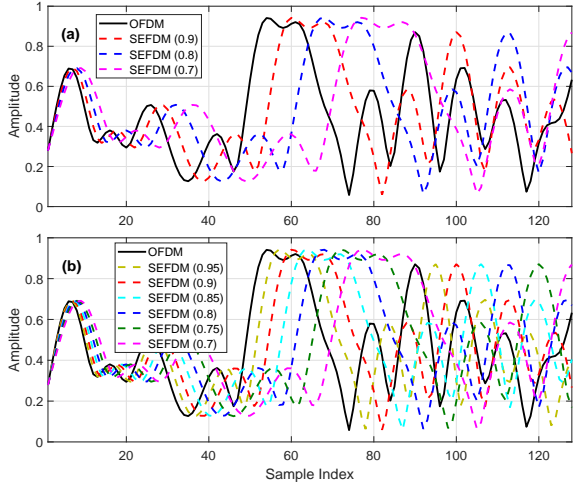}
\end{center}
\caption{Signal feature diversity and similarity visualization by modulating the same QPSK data. (a) Type-I signal pattern. (b) Type-II signal pattern.}
\label{Fig:ML_feature_combined_G_I_G_II}
\end{figure}

The time-domain SEFDM signals are illustrated in Fig. \ref{Fig:ML_feature_combined_G_I_G_II} where two types of signal patterns are presented. It is inferred that the classification of Type-I signal pattern is easier than that of Type-II signal pattern since the signal features in Type-I are more distinguishable.

The discrete format of one time-domain SEFDM symbol is defined as
\begin{equation}
X_k=\frac{1}{\sqrt{N}}\sum_{n=0}^{N-1}s_{n}\exp\left(\frac{j2{\pi}nk\alpha}N\right),\label{eq:SEFDM_discrete_signal}\end{equation}
where the expression is very similar to that of OFDM except the bandwidth compression factor $\alpha=\Delta{f}\cdotp{T}$, in which $\Delta{f}$ is the sub-carrier spacing and $T$ is the time period of one SEFDM symbol. The signal spectral bandwidth in \eqref{eq:SEFDM_discrete_signal} is compressed when $\alpha<1$ and is equivalent to that of OFDM when $\alpha=1$. The number of sub-carriers is determined by $N$. $s_{n}$ is the $n^{th}$ single-carrier symbol within one SEFDM symbol and $X_k$ is the $k^{th}$ time sample with $k=0,1,...,N-1$. The instantaneous power for one SEFDM symbol is computed in the following
\begin{equation}\label{eq:SEFDM_square_signal}
\begin{split}
|X_k|^2&=\frac{1}{N}\sum_{n=0}^{N-1}\sum_{m=0}^{N-1}s_{n}s^{*}_{m}\exp\left(\frac{j2{\pi}(n-m)k\alpha}N\right)\\
&=\frac{1}{N}\sum_{n=0}^{N-1}|s_{n}|^2+\\
&\underbrace{\frac{1}{N}\sum_{n=0}^{N-1}\sum_{m\neq{n},m=0}^{N-1}s_{n}s^{*}_{m}\exp\left(\frac{j2{\pi}(n-m)k\alpha}N\right)}_{ICI}.
\end{split}
\end{equation}

It is clearly shown that the \ac{ICI} term in \eqref{eq:SEFDM_square_signal}, which is related to the value of $\alpha$, determines the possibility of identifying different SEFDM signals. It is inferred that when SEFDM signals have similar values of $\alpha$, the ICI term will become similar and would complicate signal classification.

\section{Classification Strategies}

\subsection{CNN Classification}

\begin{figure}[t!]
\begin{center}
\includegraphics[scale=0.43]{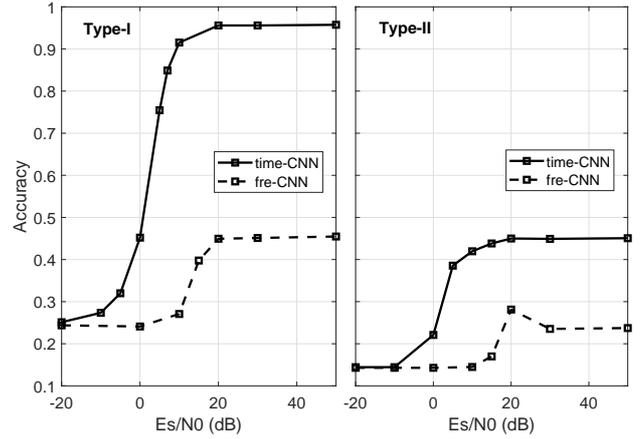}
\end{center}
\caption{CNN classification accuracy for SEFDM signals considering either time-domain or frequency-domain features.}
\label{Fig:CNN_accuracy_performance_Type_I_Type_II_wavelet}
\end{figure}

A multi-layer CNN classifier is trained in a recent work \cite{tongyang_VTC2020_DL_classification} to automatically extract signal features in either time-domain or frequency-domain. Based on extracted features, classification results are compared in Fig. \ref{Fig:CNN_accuracy_performance_Type_I_Type_II_wavelet}, in which the time-domain classifier achieves higher accuracy than its frequency-domain counterpart. Classification accuracy can reach 95\% when considering limited number of non-orthogonal signals in Type-I. However, the accuracy drops greatly when adding more similar signals in Type-II.

\subsection{SVM Classification}

The limitation of the previous work \cite{tongyang_VTC2020_DL_classification} is obvious and the motivation for this work is to accurately classify Type-II signals. The training of a multi-layer CNN classifier is time-consuming since it requires extensive hyperparameter tuning and iterative back propagation optimization. Therefore, it would be more efficient to use traditional machine learning strategies with manual feature extractions. The SVM classifier, based on domain-knowledge dependent features, is applied in this work. Firstly, the training is fast since features are obtained in advance rather than time-consuming data training. Secondly, the methodology of machine learning is deterministic and its working principle can be well explained. Since there are multiple signal classes in Type-I and Type-II, therefore a multiclass \ac{ECOC} model \cite{ECOC_Dietterich_1995} is applied here. A one-versus-one \cite{ECOC_2014_codingDesign} coding strategy is implemented for separating different classes, which simplifies the multiclass classification task into multiple binary class classification tasks. Thus, multiple binary SVM learners, with a polynomial kernel of order two, are used for the multiclass classification.

\section{Feature Selection}

This section will firstly explore the impacts of different one-dimensional statistical features and their combinations either in time-domain or frequency-domain. The second part will investigate the impact of two-dimensional time-frequency features via the single-level wavelet transform.

\subsection{Statistical Features}\label{subsec:statistical_features}

The commonly used statistical feature is arithmetic mean, which computes the average value of a dataset. Variance is used to measure the variations of a dataset. Small variance indicates that the values of dataset elements are closer to the arithmetic mean while large variance indicates that the dataset elements are spread out away from the mean. Skewness \cite{Feature_skewness_2011} is a way to measure data distribution characteristics. Negative skewness indicates that a dataset distributes more data to the left side relative to its mean; positive skewness indicates that data is more distributed to the right side of the mean. The ratio between the maximum value and the minimum value is also studied here and the MaxMin ratio can tell the fluctuations of a dataset. Interquartile range (IQR) \cite{book_IQR_1996} is a way to measure data dispersion, which equals the difference between the 25th percentile and the 75th percentile.

\subsection{Time-Frequency Features}

The previous work \cite{tongyang_VTC2020_DL_classification} revealed that independent time-domain features or frequency-domain features cannot efficiently identify Type-II signals. Therefore, the joint analysis of time-frequency signal features is important since feature diversity would be enhanced by considering two domains. This section applies wavelet transform \cite{wavelet_tour_Mallat_2008} to manually extract hidden signal features in time-frequency dimensions.

\begin{figure}[t!]
\begin{center}
\includegraphics[scale=0.25]{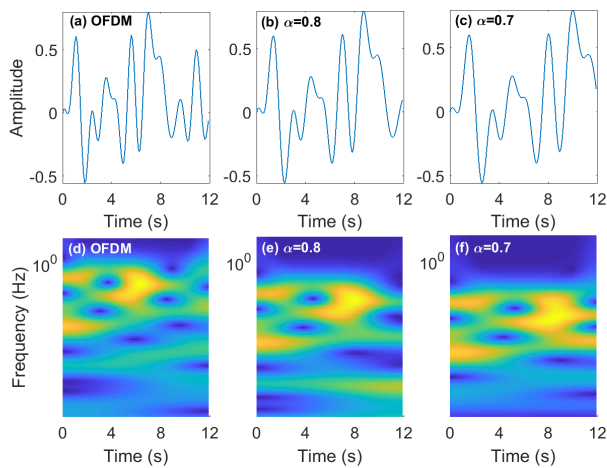}
\end{center}
\caption{Spectrogram of OFDM and SEFDM signals after wavelet transform. For the purpose of illustration, the signals are simply generated with $N$=12 sub-carriers at a data rate $R_s$=1 bit/s.}
\label{Fig:wavelet_spectrogram_1_08_07}
\end{figure}

There are two types of wavelet transform for time-frequency analysis, namely \ac{CWT} and \ac{DWT}. \ac{CWT} provides a detailed representation for signals by using fine scale factors. It therefore leads to high-resolution signal analysis and can capture crucial signal features. However, the obvious disadvantage of \ac{CWT} is its higher computational complexity over \ac{DWT}. A large time-frequency spectrogram grid would be obtained with the fine representation of scales. In this work, we would like to explore the accurate signal transient localization via detailed time-frequency analysis. Therefore, the high-resolution wavelet transform \ac{CWT} is used rather than its coarse wavelet transform \ac{DWT}.

\begin{figure}[t!]
\begin{center}
\includegraphics[scale=0.41]{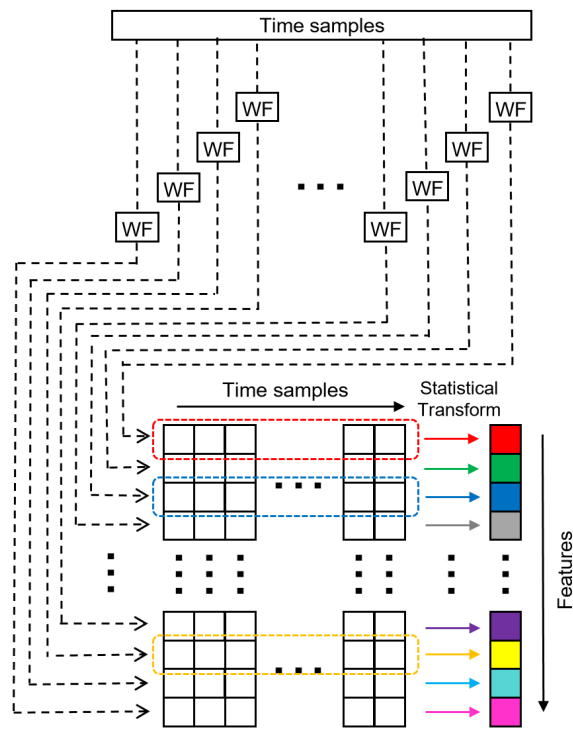}
\end{center}
\caption{ One-dimensional wavelet feature generation based on wavelet filtering and statistical feature dimensionality reduction.}
\label{Fig:time_frequency_feature_grid_transform}
\end{figure}

There are several wavelet candidates for wavelet transform. This work employs the widely used Morse wavelet and the effects of different wavelets are not taken into account. The \ac{CWT} time-frequency analysis for OFDM and SEFDM signals using Morse wavelet is illustrated in Fig. \ref{Fig:wavelet_spectrogram_1_08_07}. It is clearly shown that with the reduction of alpha, the frequency scales for SEFDM shrink to show the effect of bandwidth compression while its time scales are stretched to show the time-domain sample characteristics. Typical artificial intelligent solutions are to feed the time-frequency grid as an image to a deep learning neural network such as \ac{CNN}. However, this would cause extra training complexity since the optimal neural network hyperparameters have to be tuned based on iterative attempts. Therefore, pre-processing is required to simplify the two-dimensional time-frequency feature representation into a one-dimensional feature vector as illustrated in Fig. \ref{Fig:time_frequency_feature_grid_transform}. The strategy is to maintain the fine frequency scales of \ac{CWT} while reducing time samples dimensionality using the statistical knowledge explained in Section \ref{subsec:statistical_features}.

\section{Classifier Training and Testing}

To have a realistic training scenario, channel/hardware impairments have to be considered. The wireless channel \ac{PDP} and hardware impairments are defined in \cite{tongyang_VTC2020_DL_classification} and are reused in this work. Signals are generated according to Table \ref{tab:table_signal_specifications} where 2048 time samples are produced at the transmitter for each OFDM/SEFDM symbol. There is no synchronization mechanism between the transmitter and the receiver. Therefore, the receiver would capture 2048 time samples and randomly truncate 1024 samples for training. At the training stage, 2,000 OFDM/SEFDM symbols are generated for each class (i.e. each $\alpha$) following the data augmentation principle in \cite{tongyang_VTC2020_DL_classification}. In this case, there are overall 8,000 symbols for the Type-I signal pattern and 14,000 symbols for the Type-II signal pattern. For testing, there are overall 3,200 OFDM/SEFDM symbols for Type-I and 5,600 symbols for Type-II.

\begin{table}[t!]
\caption{Signal specifications}
\centering
\begin{tabular}{ll}
\hline \hline
$\mathbf{Parameter}$ & $\mathbf{Signal}$  \\[0.5pt] \hline 
Sampling frequency (kHz) & 200 \\ 
IFFT sample length & 2048 \\ 
Oversampling factor & 8  \\
No. of data sub-carriers & 256 \\ 
Bandwidth compression factor $\alpha$ & 1,0.95,0.9,0.85,0.8,0.75,0.7\\ 
Modulation scheme & QPSK  \\ \hline \hline
\label{tab:table_signal_specifications}
\end{tabular}
\end{table}

\begin{figure}[t!]
\begin{center}
\includegraphics[scale=0.375]{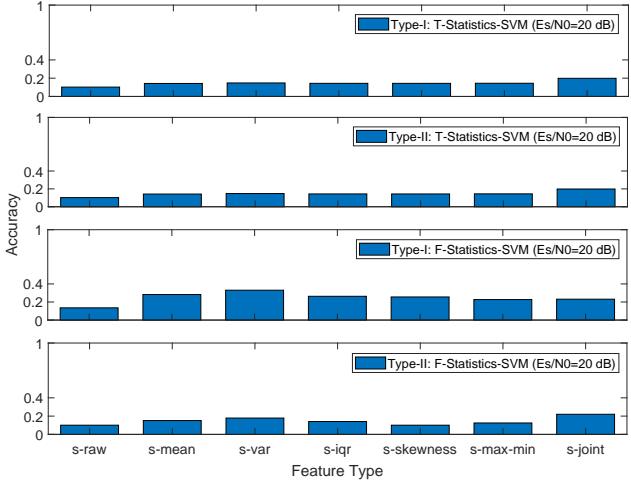}
\end{center}
\caption{Statistical feature based SVM classification accuracy trained and tested at Es/N0=20 dB. }
\label{Fig:statistic_test_singleEsN0}
\end{figure}

At first, we assume a simple training and testing scenario, in which both the training data and testing data are contaminated by \ac{AWGN} at a single Es/N0=20 dB. Multiple time-domain statistical features are extracted from the training dataset, which are labelled as `T-Statistics-SVM'. Joint statistical features are investigated by combining each statistical feature. In addition, the raw data without any manual feature extractions is also evaluated. Results in Fig. \ref{Fig:statistic_test_singleEsN0} show that all the statistical features cannot properly classify Type-I signals. It should be noted that even the joint feature cannot improve the accuracy. Similar results are obtained as well for the Type-II signals which have more challenging signal feature-similarity issues. The same feature extraction and training operations are repeated to the frequency-domain dataset, which are labelled as `F-Statistics-SVM'. The same conclusion is obtained in Fig. \ref{Fig:statistic_test_singleEsN0} that single domain statistical features cannot classify signals even in the frequency-domain.

The above results naturally lead to the joint time-frequency analysis, which would enhance the feature extraction efficiency. Wavelet transform will create a two-dimensional time-frequency feature grid. The scale range of the Morse wavelet is configured to have 7 octaves and 10 scales per octave. Therefore, considering both real and imaginary part of a signal, there are overall 140 frequency scales. In terms of time scale, following the signal specifications in Table \ref{tab:table_signal_specifications} and the 50\% random symbol truncation mechanism, 1024 time sample scales will be reserved. Therefore, \ac{CWT} will generate a pair of two-dimensional 70$\times$1024 time-frequency analysis matrices.

\begin{figure}[t!]
\begin{center}
\includegraphics[scale=0.34]{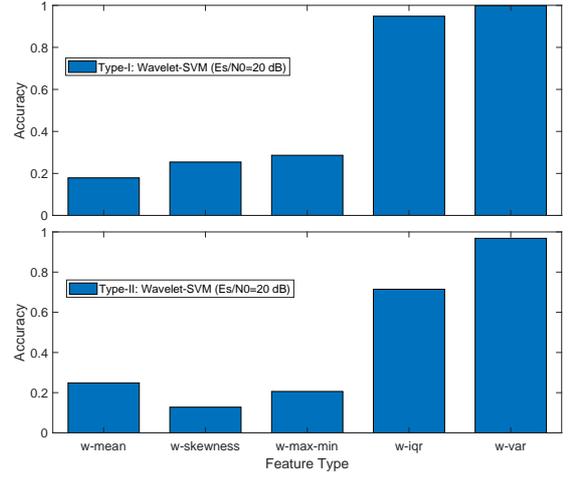}
\end{center}
\caption{Wavelet based SVM classification accuracy trained and tested at Es/N0=20 dB. }
\label{Fig:wavelet_test_singleEsN0}
\end{figure}

There are many ways to reduce the time-frequency feature dimensionality. This work applies statistical transform to reduce the amount of time samples. Thus, the two-dimensional 70$\times$1024 time-frequency grid is simplified into a 70$\times$1 frequency-scale vector following the dimensionality reduction method in Fig. \ref{Fig:time_frequency_feature_grid_transform}. Different statistical transform methods are evaluated at each frequency scale and results are shown in Fig. \ref{Fig:wavelet_test_singleEsN0}. It is clearly seen that the IQR and variance features enable higher classification accuracy than other features, which can even classify the feature-similarity dominant Type-II signals. The following classifier training will be based on those two statistical features.

\begin{figure*}[ht]
\begin{center}
\includegraphics[scale=0.48]{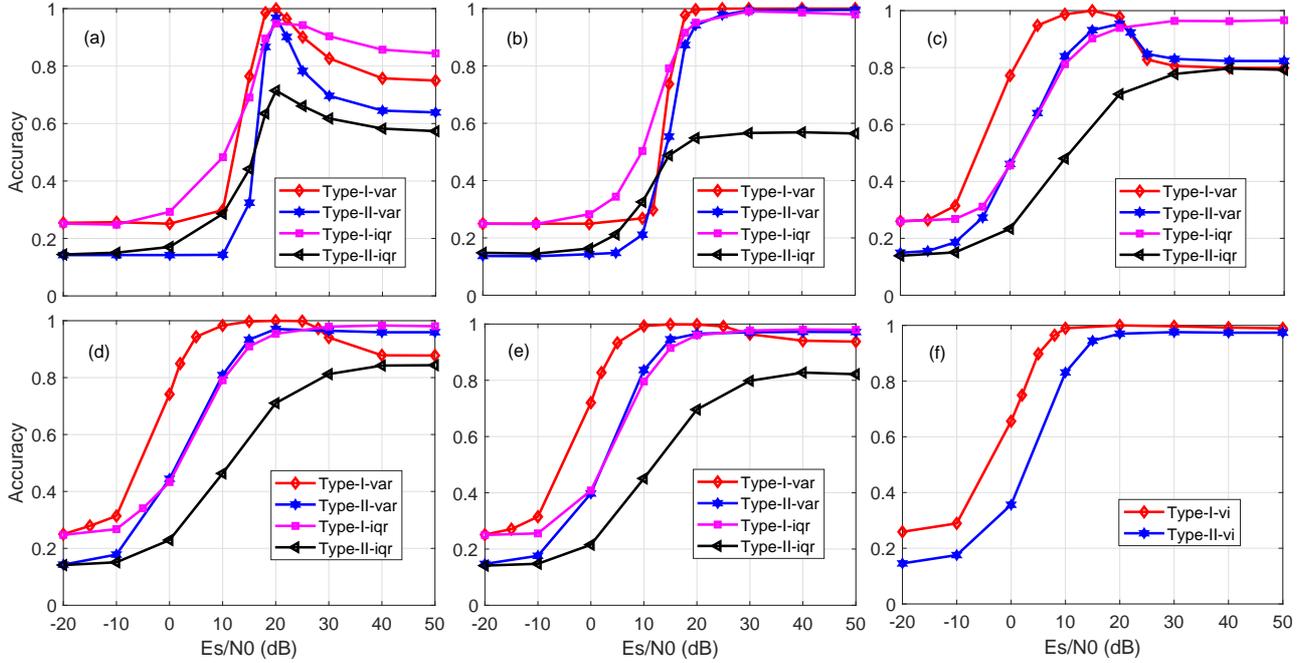}
\end{center}
\caption{Wavelet classifier accuracy tested at Es/N0 ranging from -20 dB to 50 dB. The classifier is trained at (a) Es/N0=20 dB. (b) Es/N0=20, 30, 40 dB. (c) Es/N0=0, 10, 20 dB. (d) Es/N0=0, 10, 20, 30, 40 dB. (e) Es/N0=-20, -10, 0, 10, 20, 30, 40, 50 dB. (f) Es/N0=-20, -10, 0, 10, 20, 30, 40, 50 dB. Note: var indicates variance, iqr indicates interquartile range and vi indicates variance-interquartile range.  }
\label{Fig:wavelet_combined_single_mixed_EsN0}
\end{figure*}

A wavelet classifier is firstly trained using data at a fixed Es/N0=20 dB and tested at various Es/N0 with accuracy results shown in Fig. \ref{Fig:wavelet_combined_single_mixed_EsN0}(a). It clearly shows that all the curves reach the peak accuracy at 20 dB. However, for other Es/N0 values, accuracy drops significantly. It indicates that training data at a fixed Es/N0 is not robust to train a classifier that can classify signals at a wide range of Es/N0.

To train a robust classifier, a dataset covering different Es/N0 (20, 30, 40 dB) is generated. The classification results are shown in Fig. \ref{Fig:wavelet_combined_single_mixed_EsN0}(b), in which better accuracy is reached at high Es/N0 for both Type-I and Type-II signals. However, the accuracy at low Es/N0 still needs improvement.

To enhance the classification sensitivity at low Es/N0, a dataset, covering low Es/N0 (0, 10, 20 dBs), is trained with results shown in Fig. \ref{Fig:wavelet_combined_single_mixed_EsN0}(c). All the curves are raised to achieve higher accuracy at low Es/N0. It should be noted that the variance feature enabled wavelet classifier can identify signals even below noise power and it achieves 78\% classification accuracy when Es/N0=0 dB. However, its performance drops obviously at high Es/N0, especially those beyond Es/N0=20 dB. For the IQR feature trained classifiers, both Type-I and Type-II curves are stable at high Es/N0. It should be noted that the IQR feature trained Type-I classifier outperforms the variance feature trained model at high Es/N0. It is concluded from the figure that the variance trained model is robust at low Es/N0 while the IQR trained model is robust at high Es/N0.

Based on the above results, it is inferred that classifiers trained at high Es/N0 would enable high testing accuracy merely at high Es/N0 while classifiers trained at low Es/N0 would lead to high testing accuracy at low Es/N0. This indicates that a wider Es/N0 range has to be considered for the training data. In Fig. \ref{Fig:wavelet_combined_single_mixed_EsN0}(d), classifiers are trained with data covering an Es/N0 range from 0 dB to 40 dB with an increment step of 10 dB, which basically combines the two Es/N0 ranges in Fig. \ref{Fig:wavelet_combined_single_mixed_EsN0}(b) and Fig. \ref{Fig:wavelet_combined_single_mixed_EsN0}(c). It clearly shows accuracy improvement for all the curves at both low and high Es/N0. In Fig. \ref{Fig:wavelet_combined_single_mixed_EsN0}(e), a wider Es/N0 range between -20 dB and 50 dB is considered. The variance feature trained classifier shows apparent accuracy improvement for classifying Type-I signals at high Es/N0 while all other curves have no obvious improvement. However, there is still a minor performance degradation for the variance feature based classifier at high Es/N0 when compared with the IQR trained classifier. The robust feature performance of variance at low Es/N0 and IQR at high Es/N0 inspires to combine the two features for a more reliable classifier.

The composite classifiers, trained by joint variance and IQR features, can reach high classification accuracy for both Type-I and Type-II signals at both low and high Es/N0 ranges in Fig. \ref{Fig:wavelet_combined_single_mixed_EsN0}(f). Therefore, the composite classifiers will be used in the following over-the-air experiments.

\section{Low-Cost Experiment and Results}

The experiment is operated indoor in an open space, in which facilities would cause signal reflections and further result in frequency selective channel impairments. In addition, people movement in the space would cause Doppler spread and therefore dynamic spectral fluctuations. This work will use a pair of low-cost Analog Devices \ac{SDR}
PLUTO \cite{PlutoSDR} to practically transmit and classify over-the-air signals. The signals are designed according to Table \ref{tab:table_signal_specifications} and transmitted at a free-licensed 900 MHz (33-centimeter band) carrier frequency.

\begin{figure}[t!]
\begin{center}
\includegraphics[scale=0.4]{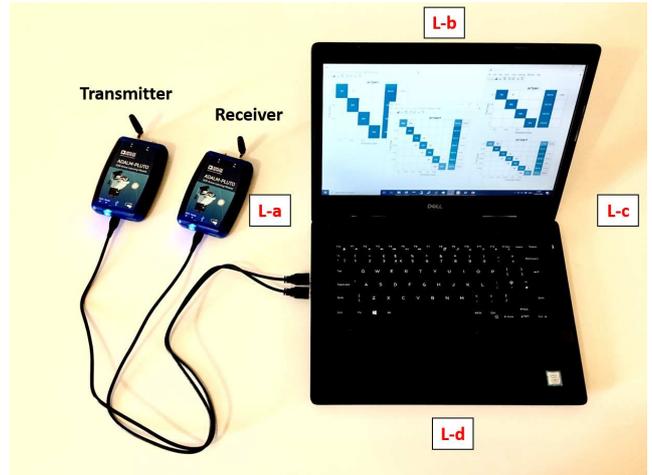}
\end{center}
\caption{Low-cost experiment setup for the wavelet classifier training and testing. Four data collection points are labelled as `L-a', `L-b', `L-c' and `L-d'.}
\label{Fig:SVM_wavelet_classification_testbed}
\end{figure}

The experiment setup, shown in Fig. \ref{Fig:SVM_wavelet_classification_testbed}, is low cost since a laptop and two PLUTO devices are sufficient to realize signal generation, over-the-air transmission, signal reception and classifier training. In order to collect diversified data from an indoor environment, we fix the position of the transmitter side SDR device and place the receiver side SDR device at different locations. In this case, a number of training datasets, impaired by channel multipath fading, power degradation and Doppler effect, are collected. Unlike the CNN classifier where a large number of training symbols are required for feature extractions, the wavelet classifier can manually extract features based on a limited dataset. Therefore, in this experiment, at each location, 400 symbols are collected for the Type-I signal pattern and 700 symbols for the Type-II signal pattern. There are four data collections considering four different locations of the receiver. Therefore, the overall collected training symbols for Type-I and Type-II are 1,600 and 2,800, respectively. For testing, the same process is repeated with four data collections. To have a fair comparison with the previous work \cite{tongyang_VTC2020_DL_classification}, the number of testing symbols per class is fixed at 800.

\begin{figure}[t!]
\begin{center}
\includegraphics[scale=0.67]{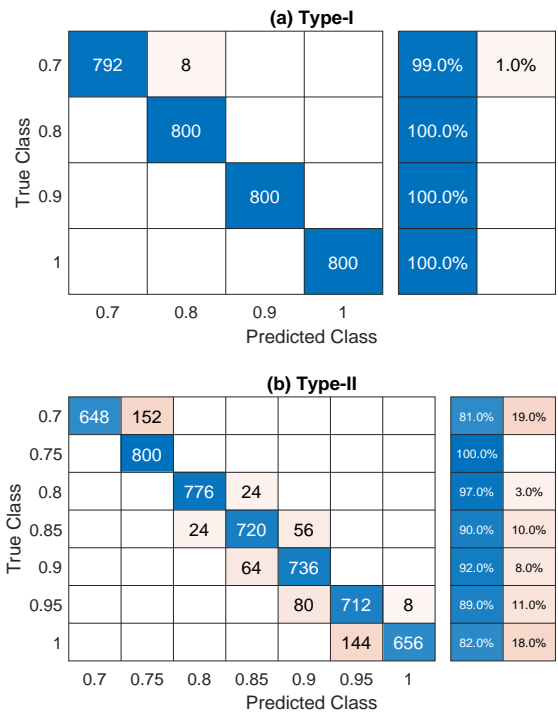}
\end{center}
\caption{Confusion matrix visualization for wavelet classification.}
\label{Fig:confusion_matrix_wavelet_classification_Type_I_Type_II}
\end{figure}

The collected data will be used to train wavelet classifiers off-line using Matlab. Once a wavelet classifier is trained, the model will be saved. Therefore, SDR devices will reuse the saved model for online signal classification and there is no need to re-train classifiers. Thus, the off-line training is a one-time operation. The confusion matrices are presented in Fig. \ref{Fig:confusion_matrix_wavelet_classification_Type_I_Type_II}. The classification accuracy for the Type-I signal pattern is nearly 100\%. For Type-II signals, the accuracy is 90\%, which is much higher than the 70.75\% in \cite{tongyang_VTC2020_DL_classification} where a transfer learning enabled \ac{CNN} classifier is applied.

\balance

\section{Conclusion}

This work aims to explore typical machine learning (ML) algorithms for non-orthogonal signal classification in non-cooperative communications. Multiple statistical approaches are tested for feature extractions in either time-domain or frequency-domain but showing unreliable classification accuracy. Wavelet transform is therefore applied to extract two-dimensional time-frequency features, which are further converted to a one-dimensional feature vector using statistical transform. Simulation results discovered that Es/N0 has great impacts on classification accuracy at the training stage. Results show increased classification accuracy over a wide range of training Es/N0. Classifiers are trained and tested with results showing that variance and IQR are the most efficient features. The combination of variance and IQR, associated with wavelet transform, enables classification accuracy up to 100\%. Furthermore, the wavelet classifier can even identify signals when the signal power is below its noise power. Results show that the variance feature enabled wavelet classifier achieves 78\% classification accuracy when Es/N0=0 dB. A low-cost experiment is set up using one laptop and two SDR devices. Practical results verify the efficacy of the wavelet enabled time-frequency features. Confusion matrices are obtained to show nearly 100\% classification accuracy for the Type-I signal pattern and 90\% accuracy for Type-II.

\bibliographystyle{IEEEtran}
\bibliography{Tongyang_Ref}

\end{document}